\documentclass[11pt]{article}

\usepackage[english]{babel}
\usepackage[latin1]{inputenc}
\usepackage{lmodern,csquotes}
\usepackage[T1]{fontenc}

\usepackage{amssymb,amsmath,amsthm,latexsym,amsfonts,amscd,dsfont,enumerate,bbm}
\usepackage{bm}
\usepackage[dvipsnames]{xcolor}
\usepackage{a4wide,graphicx,tikz}

\usepackage{algorithm,algpseudocode}
\usepackage{pgfplots}
\usepackage{pgfplotstable}
\usepackage{colortbl}
\usepackage{zref-savepos}
\pgfplotsset{compat=1.18}
\usepackage{subfig}
\usepgfplotslibrary{groupplots}
\usepackage{marginnote}
\usepackage{soul}

\usepackage[backend=biber,style=authoryear-comp,maxcitenames=1,sorting=nyt,sortcites=false]{biblatex}
\definecolor{mylightyellow}{rgb}{1,1,.8}
\definecolor{mylightgreen}{rgb}{.8,1,.8}
\definecolor{mydarkred}{RGB}{178,34,34}
\definecolor{mydarkgreen}{RGB}{34,139,34}
\definecolor{mydarkblue}{RGB}{72,61,139}
\definecolor{mydarkyellow}{RGB}{218,165,32}
\definecolor{g}{rgb}{.8,.8,.8}

\algdef{SE}[SUBALG]{Indent}{EndIndent}{}{\algorithmicend\ }%
\algtext*{Indent}
\algtext*{EndIndent}
\renewcommand{\Comment}[2][.5\linewidth]{%
  \leavevmode\hfill\makebox[#1][l]{//~#2}}

\usepackage[]{hyperref}
 \hypersetup{
 colorlinks=true,breaklinks=true,
 urlcolor= mydarkblue,linkcolor= mydarkblue,citecolor= mydarkgreen,
 pdfauthor={Manzano Herrero, A.P., Nastasi, E., Pallavicini, A., V\'azquez Cend\'on, C.},
}

\newtheorem{theorem}{Theorem}[section]
\newtheorem{remark}[theorem]{Remark}

\newcommand{\argmin}[1]{\underset{{#1}}{\operatorname{arg\,min}}\,}

\newcounter{NoTableEntry}
\renewcommand*{\theNoTableEntry}{NTE-\the\value{NoTableEntry}}

\title{Evaluating Microscopic and Macroscopic Models for Derivative Contracts on Commodity Indices \footnote{The opinions here expressed  are solely those of the authors and do not represent in any way those of their employers.}}

\pgfplotsset{
    colormap={greenred}{color=(green) color=(red)},
    nonlinear colormap trafo/.code 2 args={
        \def\max{(#2)}
        \def\min{(#1)}
        \pgfkeysalso{%
            y coord trafo/.code={%
                \pgfmathparse{##1 <= 1 ? ##1-1 : (##1-1)/(\max-1)}%
            },
            y coord inv trafo/.code={%
                \pgfmathparse{##1 <= 0 ? ##1+1 : ##1*(\max-1)+1}%
            },
        }%
    },
    nonlinear colormap around 1/.code 2 args={
        \def\max{(#2)}
        \def\min{(#1)}
        \pgfkeysalso{
            colorbar style={
                nonlinear colormap trafo={#1}{#2},
                point meta min={#1},
                point meta max={#2},
            },
            point meta={z <= 1 ? z-1 : (z-1)/(\max-1)},
        }%
    },
}

\author{
Alberto Pedro Manzano-Herrero\thanks{ Department of Mathematics and CITIC, University of A Coru{\~{n}}a, Spain {\tt alberto.manzano.herrero@udc.es}}
\and
Emanuele Nastasi\thanks{swissQuant Group AG, Zurich, Switzerland {\tt nastasi@swissquant.com}.}
\and
Andrea Pallavicini\thanks{Intesa SanPaolo Milan, Italy, {\tt andrea.pallavicini@intesasanpaolo.com}.}
\and
Carlos V\'azquez \thanks{Department of Mathematics and CITIC, University of A Coru{\~{n}}a, Spain, {\tt carlos.vazquez.cendon@udc.es}}
}
\date{
\small First Version: June 20, 2024.  This version: \today
}

\begin{document}

\maketitle

\begin{abstract}
In this article, we analyze two modeling approaches for the pricing of derivative contracts on a commodity index. The first one is a microscopic approach, where the components of the index are modeled individually, and the index price is derived from their combination. The second one is a macroscopic approach, where the index is modeled directly. While the microscopic approach offers greater flexibility, its calibration results to be more challenging, thus leading practitioners to favor the macroscopic approach. However, in the macroscopic model, the lack of explicit futures curve dynamics raises questions about its ability to accurately capture the behavior of the index and its sensitivities. In order to investigate this, we calibrate both models using derivatives of the S\&P GSCI Crude Oil excess-return index and compare their pricing and sensitivities on path-dependent options, such as autocallable contracts. This research provides insights into the suitability of macroscopic models for pricing and hedging purposes in real scenarios.
\end{abstract}

\bigskip

\noindent {\bf AMS classification codes:} 65C05, 91G20, 91G60.

\newpage
%{\small \tableofcontents}
%\vfill
%{\footnotesize \noindent The opinions here expressed  are solely those of the authors and do not represent in any way those of their employers.}
%\newpage

\maketitle
\pagestyle{myheadings} \markboth{}{{}}
\tableofcontents
\clearpage

\section{Introduction}
%%%%%%%%%%%%%%%%%%%%%%%%%%%%%%%%%%%%%%%%%%%%%%%%%%%%%%%%%
% Futures and Indices
%%%%%%%%%%%%%%%%%%%%%%%%%%%%%%%%%%%%%%%%%%%%%%%%%%%%%%%%%

Futures contracts are among the most liquid instruments in the commodities markets, making them the primary method for gaining exposure to underlying commodities. Since futures expire on specific dates, traders willing to maintain exposure to price changes in the underlying commodity must replace their expiring futures positions with new ones that have longer maturities. This process, known as a "rolling strategy" or "rolling hedge", involves shorting positions with the shortest maturities and taking long positions in futures with longer maturities. Specific implementations of rolling strategies can be found in indices like the S\&P GSCI indices\footnote{GSCI refers to Goldman Sachs Commodity Index. The white paper describing the index methodology can be found at the S\&P web site: \url{https://www.spglobal.com/spdji/en/documents/methodologies/methodology-sp-gsci.pdf}.}. Market participants seeking exposure in the commodities market can trade derivative contracts on these indices, thereby avoiding the complexities of directly trading futures, such as executing rolling strategies or managing margin requirements.\\

 %%%%%%%%%%%%%%%%%%%%%%%%%%%%%%%%%%%%%%%%%%%%%%%%%%%%%%%%%
 % Trading context
 %%%%%%%%%%%%%%%%%%%%%%%%%%%%%%%%%%%%%%%%%%%%%%%%%%%%%%%%%
In the trading desks, the typical strategy to price derivative contracts on an index starts by making a model for the index. However, when building such models, the microscopic (micro) structure, i.e. the dynamics of the underlying components of the index, is often overlooked. This can be due to several reasons. From the numerical point of view, one reason comes from the vast increase in computational demands in order to perform a separate simulation for each of the individual components. From the modeling point of view, a first reason is motivated by the difficulty in specifying coherent relations
between the components of the model. Another reason from the modeling point of view is that, although in our setting the underlying components and some derivatives contracts on them are very liquid, there are not enough liquid derivative contracts to calibrate the underlying components. The problem of not considering the micro structure is that, for certain products, the properties arising from the micro structure might have a direct impact on the prices that is not trivial to recover with a macroscopic (macro) model. Moreover, since in general the indices are not tradable assets, it is not clear how the macro models should be used for hedging purposes. \\

 %%%%%%%%%%%%%%%%%%%%%%%%%%%%%%%%%%%%%%%%%%%%%%%%%%%%%%%%%
 % Main goal
 %%%%%%%%%%%%%%%%%%%%%%%%%%%%%%%%%%%%%%%%%%%%%%%%%%%%%%%%%
The main goal of this paper is building a model for futures prices that is able to calibrate plain-vanilla option prices on both commodity futures and indices and then compare its results with that of the more standard macro model on an index. Since contracts on indices are usually sensitive to smile effects and include path-dependency, the pricing model must be able to describe both curve and smile dynamics.\\

 %%%%%%%%%%%%%%%%%%%%%%%%%%%%%%%%%%%%%%%%%%%%%%%%%%%%%%%%%
 % Literature
 %%%%%%%%%%%%%%%%%%%%%%%%%%%%%%%%%%%%%%%%%%%%%%%%%%%%%%%%%
Local volatility (LV) models, as introduced by \cite{dupire1994pricing,Derman1994}, are well-known for their ability to perfectly match the market volatility surface for plain-vanilla options, thereby accurately reproducing the volatility smile. However, LV models can sometimes be unrealistic as they tend to flatten implied forward volatilities, a limitation highlighted  in \cite{rebonato1999volatility} and \cite{hagan2002managing}, making them potentially unsuitable for pricing contracts on indices. In contrast, Stochastic Volatility (SV) models, like the Heston model (see \cite{heston1993closed}), where the variance of the asset follows its own stochastic differential equation (SDE), do not exhibit this flattening effect. Despite their advantages, SV models have a parametric form that prevents exact calibration to market prices of plain-vanilla options. In order to overcome these issues, Stochastic Local Volatility (SLV) models have been proposed, aiming to combine exact calibration to plain-vanilla options with the stochastic nature of volatility. Emerging in the late 1990s and early 2000s (for example, see \cite{said1999pricing,lipton2002masterclass,ren2007calibrating}), SLV models have since become a standard for pricing in many markets.

 %%%%%%%%%%%%%%%%%%%%%%%%%%%%%%%%%%%%%%%%%%%%%%%%%%%%%%%%%
 % Our Approach
 %%%%%%%%%%%%%%%%%%%%%%%%%%%%%%%%%%%%%%%%%%%%%%%%%%%%%%%%%
Examining our commodity settings more closely, we observe that accurately describing the dynamics of the futures term structure generally requires a distinct SV or LV model for the price of each futures contract, as noted in \cite{Pilz2011} and \cite{chiminello}. While these models provide a precise depiction of futures term structure dynamics, they can be challenging to calibrate due to the typically limited number of quotes available in commodity markets. Consequently, we seek a more parsimonious approach. Following the methodology outlined by \cite{nastasi2020smile} and \cite{commodity_indices}, we employ a smaller family of SLV processes to govern the dynamics of the entire futures term structure.\\
 %%%%%%%%%%%%%%%%%%%%%%%%%%%%%%%%%%%%%%%%%%%%%%%%%%%%%%%%%
 % Certificates
 %%%%%%%%%%%%%%%%%%%%%%%%%%%%%%%%%%%%%%%%%%%%%%%%%%%%%%%%%
In order to compare the micro and macro models, we will pay attention to the differences in prices and sensitivities of autocallable contracts, since they are path dependent. Autocallables are structured investment products which contain optionalities automatically triggered based on a specific event. These products are
triggered on predetermined observation dates if an underlying asset or reference portfolio reaches or surpasses a barrier. More precisely, on the intermediate observation dates the autocallable pays a coupon if the underlying asset is above a certain coupon barrier level and it automatically redeems if it breaches an autocall barrier level. On the last observation date, the investor receives the principal amount, or a portion thereof, along with an optional payoff on
the maturity date. The autocallability feature can be applied daily, monthly, yearly, or based on any schedule determined in accordance with the client. This feature
is frequently offered in low-yield markets, thus providing the investor with the
potential for an above-market yield, albeit with the risk of losing part of the initial capital. We refer the readers to \cite{autocallable1}, \cite{autocallable2}, \cite{autocallable3} and references therein for a more detailed description of the autocallable product. As a result of the aforementioned structure, these products are clearly path-dependent. Moreover, they are actively traded in the market, what makes them an excellent choice to study the differences between the micro and macro models in a realistic setting.\\

 %%%%%%%%%%%%%%%%%%%%%%%%%%%%%%%%%%%%%%%%%%%%%%%%%%%%%%%%%
 % Outline
 %%%%%%%%%%%%%%%%%%%%%%%%%%%%%%%%%%%%%%%%%%%%%%%%%%%%%%%%%
The paper is organized as follows. In Section \ref{sec:index} we briefly describe the strategy to replicate
a position on the index. Then, in Section \ref{sec:model_chapter2} we illustrate the modeling framework for both the micro and macro models. Next, in Section \ref{sec:calibration} we
explain the simulation and calibration procedures for both models. In Section \ref{sec:numerical_investigations} we calibrate the micro and macro models S\&P GSCI WTI Crude Oil index and investigate the differences between the micro and macro models by calculating the prices and sensitivities of different path-dependent contracts. Finally, we wrap up with the conclusions in Section \ref{sec:conclusions}.

\section{S\&P GSCI indices}
\label{sec:index}
The following section describes S\&P indices and it is identical to Section 2 of our previous work \cite{commodity_indices}.

%%%%%%%%%%%%%%%%%%%%%%%%%%%%%%%%%%%%%%%%%%%%%%%%%%%
% GSCI indices
%%%%%%%%%%%%%%%%%%%%%%%%%%%%%%%%%%%%%%%%%%%%%%%%%%

S\&P GSCI indices are designed to replicate performances of actual commodity sectors. The idea behind the construction of the S\&P GSCI indices is to create an index that simulates a continuous investment on a basket of commodities (or a single commodity). The calculation of the S\&P GSCI indices takes into account the fact that a person holding positions in a contract near expiration would need to roll such positions forward as they approach settlement or delivery dates. For this reason, the methodology for calculating S\&P GSCI indices includes a rolling procedure designed to replicate the rolling of actual futures positions. Since executing the rolling procedure on a single day could be difficult to implement or, if completed on a single day, could have an adverse impact on the market, such rolling takes place over a period of several days.\\

As the mechanics of the rolling depends particularly on which specific index we are dealing with, from now on we will focus on S\&P GSCI Excess-Return (ER) indices, which represent the daily return of a portfolio of commodity futures contracts. The rolling forward of the underlying futures contracts occurs once each month, on the fifth through ninth business day (the roll period) and the index is calculated as though these rolls occur at the end of each day during the roll period at the daily settlement prices. In the next subsections we explain the specific strategy of the non-rolling and rolling periods. As the strategy varies slightly for different indices and we will be working with the single-commodity index S\&P GSCI Crude Oil ER, we will describe how this index is calculated. This same procedure also covers how most of the single-commodity indices are built with a few exceptions (for example the S\&P GSCI Gold ER).

\subsection{Evolution during a non-rolling period}

%%%%%%%%%%%%%%%%%%%%%%%%%%%%%%%%%%%%%%%%%%%%%%%%%%%%%%
% Evolution during a non roll period
%%%%%%%%%%%%%%%%%%%%%%%%%%%%%%%%%%%%%%%%%%%%%%%%%%%%%

On any business day during a non-rolling period the value of the S\&P GSCI Crude Oil ER index is equal to the product of the value of the index on the preceding business day times one plus the contract daily return on the business day on which the calculation is made. 

For each futures maturity date $T_i, \,i=1,\ldots,M_F$, we denote $F_t(T_i)$ the futures price observed at time $t\le T_i$. Moreover, let $I_t$ denote the value of the index at time $t$. The investment strategy implemented by the index consists in buying, at the beginning of each day $t$, a quantity $Q_t$ of futures contracts on the front month such that the nominal value of the investment is exactly $I_t$. Therefore, the amount $Q_t$ of contracts we buy is given by
\begin{equation}
Q_t := \frac{I_t}{F_t(T^c)} \,,
\label{norollingQuantity}
\end{equation}
where $T^c$ indicates the maturity of the front month. Due to the market movement of the futures price, at the end of the day our investment will have generated a profit-and-loss equal to
\begin{equation}
W_{t+1} := Q_t \left( F_{t+1}(T^c) - F_t(T^c) \right) = I_t \left( \frac{F_{t+1}(T^c)}{F_t(T^c)} - 1 \right) \,.
\end{equation}
Such profit-and-loss is invested again in the strategy whose new value becomes
\begin{equation}
I_{t+1} = I_t + W_{t+1} = I_t \,\frac{F_{t+1}(T^c)}{F_t(T^c)} \,.
\end{equation}
If we repeat the same strategy the next day we get that the invested amount remains unchanged, so that
\begin{equation}
Q_{t+1} = \frac{I_{t+1}}{F_{t+1}(T^c)} = \frac{I_t}{F_t(T^c)} \,.
\end{equation}
Proceeding recursively we have that the index value at $n$-th day after $t$ can be calculated directly from the initial conditions as follows
\begin{equation}
I_{t + n} = I_t \,\frac{F_{t+n}(T^c)}{F_t(T^c)} \,.
\label{norollingStrategy}
\end{equation}

\subsection{Evolution during a rolling period}

%%%%%%%%%%%%%%%%%%%%%%%%%%%%%%%%%%%%%%%%%%%%%%%%%%%%%%%%%%%%%%%%%%%%%%%%%%%%%
% Evolution during a roll period
%%%%%%%%%%%%%%%%%%%%%%%%%%%%%%%%%%%%%%%%%%%%%%%%%%%%%%%%%%%%%%%%%%%%%%%%%%%%%

On a rolling period, we need to roll the nearest futures contract $T^c \equiv T_1$ to the second nearest futures contract $T^f\equiv T_2$ at a rate of 20\% per day for the five days of the roll period. Notice that in other indices different from the S\&P GSCI Crude Oil ER index the rolling procedure might be done between contracts with other maturities (e.g.\ the GSCI Gold ER). We could also adapt our methodology to these indices in a straightforward way.\\

Until just before the end of the fifth business day, the entire S\&P GSCI Crude Oil ER index portfolio consists of the front futures contracts. At the end of the fifth business day, the portfolio is adjusted so that 20\% of the held contracts are in the second futures contract, while 80\% remain in the front one. The roll process continues on the sixth, seventh and eighth business days, with relative weights of front to second contracts of 60\%/40\%, 40\%/60\% and 20\%/80\%. At the end of the ninth business day, the last contract of the old front futures is exchanged, thus completing the roll and leaving the entire portfolio in what we have been calling the second futures contract. At this time, this former second futures becomes the new front contract and a new second futures is formed (with futures maturities further in the future) for use in the next month roll.\\

The last key point about the roll process is to specify exactly what the 80\%/20\% or other relative splits between front and second contracts mean. The roll percentages refer to contracts or quantities, not values. Taking the first day of the roll as an example, just before the roll takes place at the end of the day, the S\&P GSCI Crude Oil ER  index consists of the front futures contract. That portfolio, constructed the night before and held throughout the fifth business day, has a dollar value. For the roll, that dollar value is distributed across the front and the second futures such that the number of contracts or the quantity of the front ones is 80\% of the total and the quantity held of the second ones is 20\% of the total.\\

We can illustrate the mechanics of the rolling period with an example to clarify the terms. Starting from the fifth business day of the month, due to the approaching expiry of the front month, the investment is gradually spread between the front and second futures. We call $\alpha(t)$ the investment percentage on the front futures contract  $F_t(T^c)$ and $1-\alpha(t)$ the investment percentage on the second futures contract $F_t(T^f)$. Then, we define $Q_t$ within the rolling period as
\begin{equation}
Q_t := \frac{I_t}{\alpha(t) F_t(T^c) + (1-\alpha(t)) F_t(T^f)} \,,
\label{rollingQuantity}
\end{equation}
which represents the purchased quantity of a fictitious contract made by the combination of the first and second futures. This investment generates the following profit-and-loss
\begin{equation}
{\rm PnL}_{t+1} := I_t \frac{\alpha(t) ( F_{t+1}(T^c) - F_t(T^c) ) + (1-\alpha(t)) ( F_{t+1}(T^f) - F_t(T^f) ) }{\alpha(t) F_t(T^c) + (1-\alpha(t)) F_t(T^f)} \,,
\end{equation}
so that the new value of the strategy is given by
\begin{equation}
\label{eq: evolution_roll}
I_{t+1} = I_t + {\rm PnL}_{t+1}  = I_t \frac{\alpha(t) F_{t+1}(T^c)  + (1-\alpha(t)) F_{t+1}(T^f) }{\alpha(t) F_t(T^c) + (1-\alpha(t)) F_t(T^f)} \,.
\end{equation}
Contrary to the non-rolling period, if we now consider the evolution to the next day we obtain that the quantities have changed due to the change in the investment weights
\begin{equation}
\begin{split}
Q_{t+1} = \frac{I_t}{\alpha(t) F_t(T^c) + (1-\alpha(t)) F_t(T^f)} \frac{\alpha(t) F_{t+1}(T^c)  + (1-\alpha(t)) F_{t+1}(T^f) }{\alpha(t+1) F_{t+1}(T^c) + (1-\alpha(t+1)) F_{t+1}(T^f) } \,,
\end{split}
\end{equation}
so that the value of the index at time $ t + 2 $ cannot be evaluated only as a function of the initial conditions.\\

The description of the index strategy leads us to introduce a futures price model able to describe the complex path-dependent dynamics of rolling periods. In the next section we deal with this problem.

\section{Modeling framework}\label{sec:model_chapter2}
The prices of a path depend options on a commodity index are expected to have a direct dependence not only on the paths followed by futures prices, but also on the correlation between futures. Hence, we are interested in the comparison of two different models. A macro model, \emph{i.e.}, a model that tries to capture the dynamics of the index regardless of its underlying microstructure and a micro model, \emph{i.e.}, a different model for each of the futures.

\subsection{Micro model}\label{sec:micro}

%%%%%%%%%%%%%%%%%%%%%%%%%%%%%%%%%%%%%%%%%%%%%%%%%%%%%%%%%%%%%%%%%%%%%%%%%%%%%%%
% Multi asset Heston model for commodities
%%%%%%%%%%%%%%%%%%%%%%%%%%%%%%%%%%%%%%%%%%%%%%%%%%%%%%%%%%%%%%%%%%%%%%%%%%%%%%%
In order to build the micro model we start by defining the dynamics of the futures prices. The index dynamics will be derived by implementing the index definition given in Section \ref{sec:index} in terms of the underlying futures. For each futures maturity date $T_i, \, i=1,\ldots,M_F$, let $F_t(T_i)$ be the futures price observed at time $t\le T_i$. We describe the dynamics of futures prices under the risk-neutral measure as in~\cite{nastasi2020smile} and we consider deterministic interest rates. In line with the work in \cite{commodity_indices}, we assume that each futures price follows a SLV model given by
\begin{equation}
dF_t(T_i) = L^F(t,T_i,F_t(T_i)) \,\sqrt{v^{F,i}_t} \,dW^{F,i}_t,
\label{eqn:SLV_model}
\end{equation}
for $i=1,\ldots,M_F$, where $\{L^F(t,T_i,K)\}_{i=1,\ldots,M_F}$ are the so-called leverage functions and represent the local-volatility components of each SLV model. The common variance processes $v^{F,i}_t$ satisfies the SDE
\begin{equation}
dv^{F,i}_t = \kappa^F (\theta^F - v^F_t) + \chi^F \sqrt{v^{v,i}_t} \,dW^{v,i}_t \,,
\label{eqn:variance}
\end{equation}
where $\kappa^F$ is the mean-reversion speed of the variance, $\theta^F$ is the long-term mean variance and $\chi^F$ is the volatility of the variance (also known as vol-of-vol). The initial value of the common variance $v_0^{F,i} = v^F_0$ is not directly observable, so that we consider it as an additional parameter to be calibrated. For simplicity, all these parameters are assumed to be constant and the same across all futures. Moreover, $\{W^{F,i}_t\}_{1,\ldots,M_F}$ and $\{W^{v,i}_t\}_{1,\ldots,M_F}$ are standard Brownian motions under the risk-neutral measure with correlations
\begin{equation}
d\langle W^{F,i},W^{F,j} \rangle_t = \rho^{F,F}_{i,j} \,dt
\;,\quad
d\langle W^{F,i},W^{v,j} \rangle_t = \delta_{i,j}\rho^{F,v} \,dt
\end{equation}
for $i,\,j=1,\ldots,M_F$. Notice that the variance processes $v_t^{F,i}$ share the same correlation value between such process and the other futures Brownian motions since we aim at a parsimonious description of futures dynamics when possible.\\

However, this model does not assume any specific form for the correlation matrix. Since the calibration of each entry in the correlation matrix would require many different contracts and a very complicated calibration strategy, we will assume that the correlation matrix follows the one parameter parametrization suggested in \cite{rebonato1999volatility}:
% http://www.packham.net/index_files/Thesis.pdf
\begin{equation}
d\langle W^{F,i},W^{F,j} \rangle_t = \rho_{i,j}^{F,F} :=  e^{-\beta |T_i-T_j|},\, i,j = 1,...,M^F,
\label{eqn:rho_chapter2}
\end{equation}
where $i,\,j=1,\ldots,M^F$ and $\beta>0$ is called the de-correlation parameter.
The assumption that the correlation matrix has the form of Equation \eqref{eqn:rho_chapter2} is somehow reasonable. On the one hand, because it always produces a valid correlation matrix in the sense that it produces a real, symmetric, positive-definite matrix. On the other hand, because in the market the correlation decreases for increasing maturity intervals and so does the proposed matrix. However, another feature of the correlation matrix that is typically observed in the markets is the fact that the correlation increases with large maturities and our model does not fulfill such restriction. In any case, for the shake of simplicity, we will restrict ourselves with correlation matrices of the form given in Equation \eqref{eqn:rho_chapter2}. 

\subsection{Macro model}\label{sec:macro}

%%%%%%%%%%%%%%%%%%%%%%%%%%%%%%%%%%%%%%%%%%%%%%%%%%%%%%%%%%%%%%%%%%%%%%%%%%%%%%%
% Macro model
%%%%%%%%%%%%%%%%%%%%%%%%%%%%%%%%%%%%%%%%%%%%%%%%%%%%%%%%%%%%%%%%%%%%%%%%%%%%%%%

The dynamics of the index that we propose resembles the one of the micro-structure. Since the index is constructed as a combination of the different futures, we propose to model the index with the same dynamics of the underlying futures
\begin{equation}
dI_t = L^I(t,I_t) \,\sqrt{v_t} I_t \,dW^I_t,
\label{eqn:SLV_single}
\end{equation}
where $L^I(t,I_t)$ is the leverage function. The variance process $v_t$ satisfies the SDE:
\begin{equation}
dv_t = \kappa^I (\theta^I - v_t) + \chi^I \sqrt{v_t} \,dW^v_t \,,
\label{eqn:variance_chapter2}
\end{equation}
Moreover, $W^I_t$ is a standard Brownian motion under the risk-neutral measure and the instantaneous correlation between the $W^I_t$ and $W^v_t$ is given by
\begin{equation}
d\langle W^I,W^v \rangle_t = \rho^{I,v} \,dt.
\end{equation}
As we can see, the dynamics of both models are very similar, the most notable differences being the futures term and the correlation structure. In the case where the initial value of the futures is the same for all maturities and the correlation structure is just the identity matrix $\rho^{F,F}_{ij} = \delta_{i,j}$, the macro model collapses to the micro model. Note that, although they have the same structure, the parmaeters in the variance processes $dv_t^i$ defined in Equation \eqref{eqn:variance_chapter2} are different from those in the variance process $dv_t$ defined in Equation \eqref{eqn:variance}.

\section{Calibration and simulation}\label{sec:calibration}
The next steps are the calibration and simulation of both models. The calibration will be different for the macro and micro models. In the micro model, we will calibrate the leverage functions to the prices of plain-vanilla options on the different futures and the correlation structure plus the stochastic parameters on plain-vanilla options on the index. In the macro model, we will calibrate the leverage function and the stochastic parameters to the prices of plain-vanilla options on the index. In the next subsections we give more details on this calibration.

\subsection{Calibration of the micro model}\label{sec:micro_calibration}
The calibration of the micro model is done in a iterative two-step process. On the one hand, the leverage function is calibrated to the prices of plain-vanilla options on the different futures by means of Markov projections via the application of the Gy\"ongy's Lemma. On the other hand, the correlation structure plus the stochastic parameters are calibrated to plain-vanilla options on the index using a global-local calibration strategy. \\ \\
\subsubsection{The leverage functions}\label{sec:leverage_calibration}

%%%%%%%%%%%%%%%%%%%%%%%%%%%%%%%%%%%%%%%%%%%%%%
% Connection with the local volatility model
%%%%%%%%%%%%%%%%%%%%%%%%%%%%%%%%%%%%%%%%%%%%%%

If we look at the SLV literature, for instance in \cite{guyon2012being}, we can find the description of a practical procedure to calibrate the leverage function by means of Markov projections via the application of the Gy\"ongy Lemma, see \cite{gyongy1986mimicking}. This Lemma states under which conditions the marginal densities of two semi-martingales are equivalent in law. Thus, if we are able to calibrate plain-vanilla options quoted by the market by means of a simpler model, e.g.\ a LV model, we can ensure that the SLV model does the same if such models satisfy the hypotheses of the Lemma.

In our case, the market quotes plain-vanilla options of futures prices, so that we introduce a LV model for each futures price, which is given by
\begin{equation}
d{\hat F}_t(T_i) = \hat{L}^F(t,T_i,{\hat F}_t(T_i)) \,d{\hat W}^{F,i}_t \,,
\label{eqn:futures_lv_equation}
\end{equation}
where the local-volatility functions $\{\hat{L}^F(t,T_i,K)\}_{i=1,\ldots,M_F}$ are assumed to be Lipschitz, positive and at most of linear growth in price, so that there exists a solution for the SDE. Then, we can apply the Gy\"ongy's Lemma to perform the matching of the marginal distributions and we obtain
\begin{equation}
\hat{L}^F(t,T_i,K) = L^F(t,T_i,K) \,\sqrt{\mathbb{E}[v^{F,i}_t|F_t(T_i)=K]} \,.
\label{eqn:leverage}
\end{equation}
We can solve for the leverage functions to write the SLV model in terms of the local volatilities, thus getting
\begin{equation}
dF_t(T_i) = \hat{L}^F(t,T_i,F_t(T_i)) \sqrt{ \dfrac{v^{F,i}_t}{\mathbb{E}[v^{F,i}_t|F_t(T_i)]} } \,dW^i_t \,.
\label{eqn:futures_master_equation}
\end{equation}

\begin{remark}\label{remark:leverage1}
We notice that the existence of solutions of McKean SDEs such as the one in Equation \eqref{eqn:futures_master_equation} is a known open problem (see \cite{guyon2012being} for instance) and falls outside
the scope of this paper. 
\end{remark}

The problem of calibrating the leverage functions has been transformed into the simpler problem of calibrating the local volatilities $\hat{L}^F$. This problem is considered and solved in a parsimonious way in \cite{nastasi2020smile}. Here, we adopt such solution, that we briefly describe in the following paragraphs.

First, we introduce a common driving factor $\hat{s}_t$ for all future prices. We can understand it as a normalised ``spot'' price. The dynamics of $\hat{s}_t$ is given by
\begin{equation}
d\hat{s}_t = a (1 - \hat{s}_t) \,dt + \hat{s}_t \hat{L}^s(t,\hat{s}_t) \,d\hat{W}^{s}_t \,,
\end{equation}
where $a$ is the mean-reversion speed of the spot process, $\hat{L}^s(t,K)$ is the local-volatility function of the spot price. This function is assumed to be Lipschitz, positive and bounded in price. Moreover, $\hat{W}^{\hat{s}}_t$ is a standard Brownian motion under the risk-neutral measure. Then, we assume that futures prices in the LV model can be calculated starting from the normalised spot price as
\begin{equation}
{\hat F}_t(T_i) := F_0(T_i) \,\mathbb{E}_t[s_{T_i}] = F_0(T_i) \left(1 - (1-\hat{s}_t) \, e^{-a(T_i-t)}\right)
\label{eqn:original_master_relation}
\end{equation} 
for $i=1,\ldots,M_F$, where $F_0(T_i)$ is the term structure of futures prices observed in the market. Last equality in \eqref{eqn:original_master_relation} can be derived by straightforward algebra due to the particular form of the spot price dynamics. Furthermore, we can apply the It\^o Lemma and compare the result with equation \eqref{eqn:futures_master_equation}, so that we get
\begin{equation}
\hat{L}^F(t,T_i,K) := \left( K - F_0(T_i) \left( 1 - e^{-a(T_i-t)} \right) \right) \hat{L}^s(t,k^F(t,T_i,K)) \,,
\label{eq:etaF}
\end{equation}%
where the effective strike $k_F$ can be defined as
\begin{equation}
k^F(t,T_i,K) := 1 - e^{a(T_i-t)} \left( 1 - \frac{K}{F_0(T_i)} \right) \,.
\label{eq:effK}
\end{equation}%

Thus, the problem is now to calibrate the local volatility $\hat{L}^s(t,K)$ of the spot price for a given mean-reversion speed $a$ to the prices of plain-vanilla options on futures. Notice that the problem is now much simpler since instead of calibrating a different function for each futures we need to calibrate only one function. In order to do so, we write the price $C_0^F(T_i^{\text{pv}},T_i,K)$ of plain-vanilla options with expiry $T^{\text{pv}}_i$ on futures with maturity $T_i>T^{\text{pv}}_i$ and strike $K$ in terms of the spot price (in order to simplify notation, we assume that expiry and payment of the options are on the same date):
\begin{equation}
C_0^F(T^{\text{pv}}_i,T_i,K) := P_0(T^{\text{pv}}_i) F_0(T_i) e^{-a(T_i-t)} c^F(T^{\text{pv}}_i,k^F(T^{\text{pv}}_i,T_i,K)) \,,
\label{eqn:call_price}
\end{equation}
where $P_0(T^{\text{pv}}_i)$ is the zero-coupon bond with maturity $T^{\text{pv}}_i$ and the normalized call prices $c^F(t,k)$ are defined as
\begin{equation}
c^F(t,k) := \mathbb{E}_{0}[(\hat{s}_t-k)^+]
\end{equation}
and they satisfy the following extended version of the Dupire equation, see \cite{nastasi2020smile}:
\begin{equation}
\partial _tc^F(t,k) = \left(-a-a(1-k)\partial_k+\dfrac{1}{2}k^2\hat{L}^s (t,k)^2\partial^2 _k\right)c^F(t,k) \,.
\label{eqn:plain_vanilla_PDE}
\end{equation}
In order to calibrate $\hat{L}^s(t,k)$ we follow the same iterative strategy as in \cite{nastasi2020smile}. This strategy can be summarised in four main steps:
\begin{enumerate}
    \item Solve Equation \eqref{eqn:plain_vanilla_PDE} for a fixed $\hat{L}^s(t,k)$. We use a Cranck-Nicholson scheme in time and central finite differences in space (see \cite{wilmott1995mathematics}).
    \item Compute the Black-Scholes volatilities $ \hat{\sigma}^F$ from the model prices $c^F$.
    \item Compare the model volatilities $\hat{\sigma}^F$ with the market volatilities $\sigma^F$ and update the $\hat{L}^s(t,k)$ accordingly.
    \item Go back to step one until the target accuracy is met.
\end{enumerate}

Once $\hat{L}^s(t,k)$ is known we can calculate $\hat{L}^F(t,T_i,K)$ by means of equation \eqref{eq:etaF} and then we get the leverage functions $L^F(t,T_i,K)$ by solving equation \eqref{eqn:leverage}.\\ \\
\subsubsection{Stochastic parameters and correlation structure}\label{sec:global_local_calibration}
In Section \ref{sec:leverage_calibration} we already described how to calibrate the leverage functions $L^F(t,T_i,K)$ to recover the prices of plain vanillas of the futures. The remaining degrees of freedom of the micro model:
\begin{equation}
    \bm{p}^F = \{a,\beta,\kappa^F,\theta^F,\chi^F,\rho^{F,v},v^F_{0}\} \,,
\end{equation}
are adjusted to recover the implied volatilities of plain vanillas on the on GSCI ER index $\bm{\sigma}^I_{\rm market}$. For this purpose, we define the cost function:

\begin{equation}\label{eqn:loss_function_micro}
    D_{\ell^1}(\bm{p}^F) = D_{\ell^1}(\bm{\sigma}^I_{\rm market},\bm{\sigma}^I_{\rm micro}(\bm{p}^F)) = \sum _{j = 1}^J\left|\sigma^{I,j}_{\rm market}-{\sigma}^{I,j}_{\rm micro}(\mathbf{p}^F)\right| \,,
\end{equation}
where $\bm{\sigma}^I_{\rm micro}$ are the implied volatilities given by the micro model when evaluated with parameters $\bm{p^F}$ and $J$ is the total number of vanillas on the index that have been considered.\\

In this setup, the calibration of the free parameters to fit the prices of plain-vanillas on the index can be formulated as the following unconstrained global optimization problem in a bounded domain:
\begin{equation}\label{eqn:minimization_problem2}
    \min_{\mathbf{p}^F\in P\subseteq \mathbb{R}^7}D_{\ell^1}(\mathbf{p}^F) \ ,
\end{equation}
where $D_d$ is the cost function defined on $P^F = \Pi_{r = 1}^7[l^F_r,u^F_r]$, with $l^F_r$ and $u^F_r$ being the lower and upper bounds in direction $i$, respectively. The solution vector $\mathbf{p}_*^F$ contains the calibrated parameters and is defined as:
\begin{equation}\label{eqn:parameters_solution_micro}
        \mathbf{p}_*^F = \argmin{{\mathbf{p}^F\in P^F\subseteq \mathbb{R}^7}}D_{\ell^1}(\mathbf{p}^F),
\end{equation}
Note that each evaluation of the cost function requires the numerical solution of the micro model.\\

In order to solve the unconstrained global optimization problem (\ref{eqn:minimization_problem2}), we propose a simplified version of a two-phase calibration strategy (see \cite{FERREIROFERREIRO2020467, two_phase} for a complete description). In particular, we start by defining an initial guess $\mathbf{p}_0^F$ of the optimal solution $\mathbf{p}_*^F$. We typically choose $\mathbf{p}_0^F$ randomly. Next, we run a global optimization algorithm starting with $\mathbf{p}_0^F$. Once the global algorithm has finished, it provides an intermediate solution $\mathbf{p}_1^F$. Then, we use $\mathbf{p}_1^F$ as the initial point for a local optimization algorithm. Finally, the local optimization algorithm gives us $\mathbf{p}_2^F$. These steps are summarized in Algorithm \ref{alg:overall}.\\

\begin{algorithm}[hbtp!]
\caption{Overall optimization algorithm}\label{alg:overall}
\begin{algorithmic}
\State \textbf{Input:}
\Indent
\State $\mathbf{p}_0^F$ \Comment{random seed for $\mathbf{p}^F$}
\State $D_d$ \Comment{function to be minimized}
\EndIndent
\State \textbf{Output:}
\Indent
\State $\mathbf{p}_2^F$ \Comment{approximated value of $\mathbf{p^F_*}$}
\EndIndent
\State \textbf{Algorithm:}
\Indent
\State $\mathbf{\check p}^s_1 = \text{ESCH}(\mathbf{\check p}^s_0)$ \Comment{global minimization.}
\State $\mathbf{\check p}^s_2 = \text{Subplex}(\mathbf{\check p}^s_1)$ \Comment{local minimization.}
\EndIndent
\end{algorithmic}
\end{algorithm}
As the global optimization algorithm, we use the so-called ESCH (see \cite{DE_variation}). ESCH algorithm belongs to the broad category of Evolutionary Algorithms (EAs) (see \cite{EA}) which is a class of heuristics inspired by natural selection in biological populations. We specifically use ESCH algorithm because it is available in the \textit{NLopt} nonlinear-optimization package (see \cite{NLopt} for more details). In Algorithm \ref{alg:differential_evolution} we show the pseudocode for the ESCH method.

This evolutionary algorithm starts by dividing the population into two groups: parents and offspring. First, the parameters of the parents are randomly initialized from a uniform distribution. Then, some of the parents are selected and their information recombined to generate offspring. Some of the offspring would mutate one of their parameters. Next, we assign a score to the offspring based on their value in the function that we want to minimize. Finally, all individuals are ranked according to their fit score, the less fit individuals are
removed from the population and only the best individuals are
stored along the generations. The adopted recombination and
the mutation operators are the so-called single point and the
Cauchy distribution, respectively. The scheme for the evolution can be found in Algorithm \ref{alg:differential_evolution}. For a more detailed description we refer to the original article \cite{DE_variation}.\\

The advantage of starting by a global optimization algorithm is that it is able to escape from local minima. However, in general these algorithms in are computationally very expensive as they have a slow rate of convergence. Our purpose starting with a global optimization algorithm is to explore the space and end up with a good initial point for the local optimization algorithm. In the best case scenario the output of the algorithm should lay in the convex region defined by the global minima. Other possible choices for the global optimization routine are: Simulated Annealing (SA, see \cite{EA}, \cite{aarts1985statistical}), Differential Evolution (DE, see \cite{storn1997differential}) or Particle Swarm (PS, see \cite{kennedy1995particle}).\\
 
\begin{algorithm}[hbtp!]
\caption{ESCH pseudocode}\label{alg:differential_evolution}
\begin{algorithmic}
\State \textbf{Input:}
\Indent
\State $R$ \Comment{problem dimension}
\State $D_d$ \Comment{function to be minimized}
\State $x_0$ \Comment{initial approximation to minimum}
\State $np$ \Comment{number of parents}
\State $no$ \Comment{number of offspring}
\EndIndent
\State \textbf{Output:}
\Indent
\State $x$ \Comment{computed minimum}
\EndIndent
\State \textbf{Algorithm:}
\Indent
\State Initialize parents and offspring population
\State Parents fitness evaluation
\While{termination test not satisfied}
\State Crossover
\State Gaussian Mutation
\State Offspring fitness evaluation
\State Selection of the fittest as parents
\EndWhile
\EndIndent
\end{algorithmic}
\end{algorithm}
 As the local optimization algorithm, we use a variation of the Nelder-Mead algorithm (see \cite{Nelder1965ASM}), called ``Subplex'' developed in \cite{subplex}. The Subplex algorithm starts by dividing the search space into subspaces. Then, the Nelder-Mead algorithm is used to minimize in each subspace. The subspaces in which the minimization has been larger are joined to form a new subspace. The process continues iteratively until the stopping criteria is met. An outline of the Subplex basic steps can be found in Algorithm \ref{alg:subplex}. For a more detailed description we refer to the original thesis \cite{subplex}.\\ 
 
Note that local search algorithms are not able to escape from local minima, although they exhibit faster convergence rates than the global optimization ones. Therefore, the purpose of using the local optimization algorithm is to perform a fine grain optimization much faster than it would be possible with a global optimization method.

\begin{algorithm}[hbtp!]
\caption{Subplex pseudocode}\label{alg:subplex}
\begin{algorithmic}
\State \textbf{Input:}
\Indent
\State $R$ \Comment{problem dimension}
\State $D_d$ \Comment{function to be minimized}
\State $x_0$ \Comment{initial approximation to minimum}
\State $Scale$ \Comment{initial step size for the $r$ coordinate directions}
\State $\alpha$ \Comment{reflection coefficient}
\State $\beta$ \Comment{contraction coefficient}
\State $\gamma$ \Comment{expansion coefficient}
\State $\delta$ \Comment{shrinkage coefficient}
\State $\psi$ \Comment{simplex reduction coefficient}
\State $\Omega$ \Comment{step reduction coefficient}
\State $nsmin$ \Comment{minimum subspace dimension}
\State $nsmax$ \Comment{maximum subspace dimension}
\EndIndent
\State \textbf{Output:}
\Indent
\State $x$ \Comment{computed minimum}
\EndIndent
\State \textbf{Algorithm:}
\Indent
\While{termination test not satisfied}
\State Set stepsizes
\State Set subspaces
\For{each subspace}
\State Use Nelder Mead Simplex to search subspace
\State Check Termination
\EndFor
\EndWhile
\EndIndent
\end{algorithmic}
\end{algorithm}

\subsection{Calibration of the macro model}\label{sec:macro_calibration}
The calibration of the macro model will be different from the calibration of the micro model, since both the leverage function and the stochastic parameters need to be calibrated on the same surface of plain vanillas on the index. In a trading desk, these two are typically calibrated separately. First they build a local volatility model and calibrate it to the surface of plain vanillas on the index. Second, they calibrate a pure stochastic model on the volatility smile for a fixed maturity in the same surface. Finally, they calibrate the leverage function by means of Markov projections via the application of the Gy\"ongy's Lemma.\\ \\
In our case there are two main differences. First, the calibration of the stochastic parameters is done on the whole surface of plain vanillas on the index and not on a fixed maturity. Second, the calibration is not performed on the market surface but on the surface produced by the micro model. In this way, we ensure that the marginal probabilities of both models are the same.
\subsubsection{The leverage function}
The procedure to calibrate the leverage function for the index model follows the same steps as in Section \ref{sec:leverage_calibration}. First, we define the local-volatility model:
\begin{equation}\label{eq:index_local_vol}
    d\hat{I}_t = \hat{L}^I (t,\hat{I}_t) \hat{I}_t \, dW^{\hat{I}}_t \, .
\end{equation}
where the local-volatility $\hat{L}_I(t,K)$ is assumed to be Lipschitz, positive and bounded, so that there exists a solution for the SDE. Next, applying Gy\"ongy's Lemma we match the marginal distributions of the local-volatility model defined in Equation \eqref{eq:index_local_vol} with the SLV model defined in Equation \eqref{eqn:SLV_single}:
\begin{equation}\label{eqn:index_master_equation}
    L_I(t,K) = \dfrac{\hat{L}^I (t,K)}{\sqrt{\mathbb{E}[v_t|I_t = K]}}\,.
\end{equation}
\begin{remark}
As in Remark \ref{remark:leverage1} we stress that the existence of solutions of McKean SDEs such as the one in Equation \eqref{eqn:index_master_equation} is a known open problem (see \cite{guyon2012being} for instance) and falls outside
the scope of this paper. 
\end{remark}
Now, we can solve for the leverage function to write the SLV model in terms of the local volatilities, thus getting:
\begin{equation}\label{eq:index_slv_projected}
    dI_t = \hat{L}^I(t,I_t) I_t\sqrt{\dfrac{v_t}{\mathbb{E}[v_t|I_t]}}\, dW^I_t \,.
\end{equation}
With these manipulations, the problem of calibrating the leverage has been transformed in that of calibrating the local-volatility function $\eta^I(t,K)$. This local-volatility function satisfies a Dupire equation:
\begin{equation}\label{eq:dupire}
    \partial_t c^I(t,k) = \dfrac{1}{2}k^2 \hat{L}^I(t,k)^2 \partial_{kk} \, c^I(t,k) \,,
\end{equation}
where $c^I$ are normalised call prices defined by:
\begin{equation}
    c^I(t,k) :=\dfrac{1}{I_0} \mathbb{E}_t[(I_t-K)^+]\,,    
\end{equation}

In order to calibrate $\hat{L}^I(t,k)$ we follow an analogous strategy as in Section \ref{sec:leverage_calibration}. This strategy can be again summarised in four main steps:
\begin{enumerate}
    \item Solve Equation \eqref{eq:dupire} for a fixed $\hat{L}^I(t,k)$. We use a Cranck-Nicholson scheme in time and central finite differences in space \cite{wilmott1995mathematics}.
    \item Compute the Black-Scholes volatilities $\hat{\sigma}^I_{\text{macro}}$ from the model prices $c^I$.
    \item Compare the model volatilities $\hat{\sigma}^I_{\text{macro}}$ with the micro volatilities $\sigma^I_{\text{micro}}$ and update the local-volatility function $\hat{L}^I(t,k)$ accordingly.
    \item Go back to step one until the target accuracy is met.
\end{enumerate}
\subsubsection{The stochastic volatility}
In the micro model we calibrated the leverage and the stochastic parameters in a different set of options. However, since usually we don't have access to additional liquid options, for the macro model the stochastic parameters are typically calibrated on the same set of options as the leverage. For this purpose, we define the pure stochastic volatility model:
\begin{equation}\label{eq:sv_model}
    d \overline{I}_t = \overline{I}_t \sqrt{\overline{v}}_t d\overline{W}^{I}_t\,,
\end{equation}
where the variance process $\overline{v}_t$ satisfies the SDE:
\begin{equation}
d\overline{v}_t = \overline{\kappa}^I (\overline{\theta}^I - \overline{v}_t) + \overline{\chi}^I \sqrt{\overline{v}_t} \,d\overline{W}^{v,I}_t \,,
\label{eqn:variance_chapter2_pure_stochastic}
\end{equation}
 and the instantaneous correlation between the $\overline{W}^{I}_t$ and $\overline{W}^{v,I}_t$ is given by
\begin{equation}
d\langle \overline{W}^{I},\overline{W}^{I,v} \rangle_t = \overline{\rho}^{I,v} \,dt \,.
\end{equation}
This model has five degrees of freedom:
\begin{equation}
    \mathbf{\overline{p}}^I := \{\overline{\kappa}^I,\overline{\theta}^I,\overline{\chi}^I,\overline{\rho}^{I,v},\overline{v}^I_0\} \,.
\end{equation}
that will be calibrated on the implied volatilities of plain vanillas on the index given by the micro model $\bm{\sigma}^I_{\rm micro}$. For this purpose, we define the cost function:
\begin{equation}\label{eqn:loss_function_macro}
    D_{\ell^1}(\bm{p}^I) = D_{\ell^1}(\bm{\sigma}^I_{\rm micro},\overline{\bm{\sigma}}^I_{\rm macro}(\bm{\overline p}^I)) = \sum _{j = 1}^J\left|\sigma^{I,j}_{\rm micro}-\overline{{\sigma}}^{I,j}_{\rm macro}(\mathbf{\overline p}^I)\right| \,,
\end{equation}
where $\overline{\bm{\sigma}}^I_{\rm macro}$ are the implied volatilities given by the pure stochastic macro model and $J$ is the total number of vanillas on the index that have been considered.\\

This setup closely resembles the one in Section \ref{sec:global_local_calibration} and the problem can be formulated as the following unconstrained global optimization problem in a bounded domain:
\begin{equation}\label{eqn:minimization_problem_chapter22}
    \min_{\mathbf{\overline{p}^I}\in \overline{P}^I\subseteq \mathbb{R}^5}D_{\ell^1}(\mathbf{\overline{p}^I}) \,,
\end{equation}
where $D_d$ is the cost function defined on $\overline{P}^I = \Pi_{r = 1}^5[\overline{l}^I_r,\overline{u}^I_r]$, with $\overline{l}^I_r$ and $\overline{u}^I_r$ being the lower and upper bounds in direction $r$, respectively. The solution vector $\mathbf{\overline{p}}_*^I$ contains the calibrated parameters and is defined as:
\begin{equation}
        \mathbf{\overline{p}_*}^I = \argmin{{\mathbf{\overline{p}}^I\in \overline{P}\subseteq \mathbb{R}^5}}D_{\ell^1}(\mathbf{\overline{p}}^I),
\end{equation}
Once again, in order to solve this unconstrained global optimization problem, we replicate the same hybrid global-local calibration as in Section \ref{sec:global_local_calibration}, i.e., we start with a random seed $\overline{p}^I_0$. Next, we first apply the ESCH global minimization algorithm to obtain an intermediate result $\overline{p}^I_1$. Finally, we apply the Sublplex local minimization algorithm obtaining the solution $\overline{p}^I_2$.\\ 

With the solution of the pure stochastic volatility model $\overline{p}^I$, we calibrate the macro model by simply fixing the values the stochastic parameters of $I_t$ with the values of $\overline{I}_t$:
\begin{equation}
    \kappa^I = \overline{\kappa}^I\,, \theta^I = \overline{\theta}^I\,, \chi^I = \overline{\chi}^I\,, \rho^{I,v} = \overline{\rho}^{I,v}\,, v_{0}^I = \overline{v}_0^I.
\end{equation}

\subsection{Simulation scheme for the micro and macro models}
\label{sec:simulation_micro_macro}

%%%%%%%%%%%%%%%%%%%%%%%%%%%%%%%%%%%%%%%%%%%%%%%%%%%%%%%%%%%%%%%%%%%%%%%%%%%%
% Simulation scheme
%%%%%%%%%%%%%%%%%%%%%%%%%%%%%%%%%%%%%%%%%%%%%%%%%%%%%%%%%%%%%%%%%%%%%%%%%%%%
For the simulation of both the micro and the macro model we resort to a Monte Carlo simulation since it is flexible, simple and efficient. We will focus on three main parts. First, the simulation of the conditional expectation in the diffusive term for both the micro and the macro model. Second the discretization of the variance process. Third, the simulation of the futures and index processes.\\

The simulation of the conditional expectation in the diffusive term for both the micro and the macro is achieved by means of the particle method described in \cite{guyon2012being}. In the case where we have $N$ sample paths $\{(F_t(T_i)^{n},v^{i,n}_t)\}_{n = 1,\ldots,N}$ for the $i$-th future in the case where we have $N$ sample paths $\{(I_t^{n},v^{n}_t)\}_{n = 1,\ldots,N}$ for the macro model, the conditional expectations are approximated as:
\begin{equation}
\dfrac{1}{\sqrt{\mathbb{E}[v^i_t|F_t(T_i)]}} = \sqrt{\dfrac{\sum_{j=1}^N \delta^{\epsilon}({F}_t(T_i)^{n}-{F}_t(T_i)^{j}) }{ \sum_{j=1}^N {v}^{i,j}_t \,\delta^{\epsilon}({F}_t(T_i)^{n}-{F}_t(T_i)^{j})}}\,, 
\end{equation}%
and
\begin{equation}
\dfrac{1}{\sqrt{\mathbb{E}[v_t|I_t]}} = \sqrt{\dfrac{\sum_{j=1}^N \delta({I}_t^{n}-{I}_t^{j}) }{ \sum_{j=1}^N {v}^{j}_t \,\delta({I}_t^{n}-{I}_t^{j})}}\,. 
\end{equation}%
where the function $\delta^\epsilon(x)$ is any mollifier of the Dirac delta.\\

The simulation of the variance processes in both models is achieved through the full-truncated scheme as described in \cite{lord2010comparison}. When we have a time grid $\{t_m\}_{m=1,\ldots,M}$ the full-truncated  scheme applied to the micro model yields:
\begin{equation}
    {v}^{i,n}_{t_{m+1}} = {v}^{i,n}_{t_m} + { \kappa}^F ( { \theta}^F - ({ v}^{i,n}_{t_m})^+ ) \,\Delta t_m + { \chi}^F \sqrt{ ( { v}^{i,n}_{t_m} )^+ } \,\Delta { W}^{v,i,n}_{t_m}\,,
\end{equation}
and to the macro model:
\begin{equation}
    {v}_{t_{m+1}} = {v}_{t_m} + { \kappa}^I ( { \theta}^I - ({ v}_{t_m})^+ ) \,\Delta t_m + { \chi}^I \sqrt{ ( { v}_{t_m} )^+ } \,\Delta { W}^{v}_{t_m}\,.
\end{equation}
The  full-truncated scheme avoids the negative values of the variance process at the discrete level while maintaining a good convergence to the continuous process.\\

Finally, for the discretization of the futures and the index we follow a standard Euler-Maruyama scheme. For the micro model yields
\begin{equation}
F_{t_{m+1}}(T_i)^{n} = F_{t_{m}}(T_i)^{n} + L^F(t_m,T_i,F_{t_{m}}(T_i)^{n}) \,\sqrt{ \frac{ ({ v}^{i,n}_{t_m})^+ \,\sum_{j=1}^N \delta^\epsilon(F_{t_{m}}(T_i)^{n}-F_{t_{m}}(T_i)^{j}) }{ \sum_{j=1}^N ({v}^{i,j}_{t_m})^+ \,\delta^\epsilon(F_{t_{m}}(T_i)^{n}-F_{t_{m}}(T_i)^{j}) } } \,\Delta W^{i,n}_{t_m}\,,
\end{equation}
and for the macro model
\begin{equation}
I_{t_{m+1}}^n = I_{t_{m}}^n+I_{t_{m}}^n\sqrt{v_{t_m}^n\dfrac{\sum_{j=1}^N \delta({I}_{t_m}^{n}-{I}_{t_m}^{j}) }{ \sum_{j=1}^N {v}^{j}_{t_m} \,\delta({I}_{t_m}^{n}-{I}_{t_m}^{j})}}dW^I_{t_m}\,.
\end{equation}
The local volatility functions are obtained by making a piecewise-constant interpolation in time and a piecewise linear interpolation in price.

\section{Numerical investigations}\label{sec:numerical_investigations}
In this section we investigate the impact that the use of the two different models has on the prices and sensitivities of path-dependent options on an index. More precisely, we focus on derivatives contracts on the S\&P GSCI Crude Oil Index Excess Return composed by WTI Crude Oil futures. 
\subsection{Calibration}
As explained in Section \ref{sec:calibration} the calibration of the micro model is performed against plain vanillas on the futures and the index quoted the 16th of December of 2019. Since not all parameters are relevant for the calibration we keep some of them fixed. For the micro model we fix the parameters in Table \ref{tab:fixed_values_micro}.
\begin{table}[H]
    \centering
    \begin{tabular}{|c|c|}
        \hline
         $\bm{p}^F$ & \textbf{Value}  \\
         \hline
         $\kappa^F$ & $1.0$\\
         \hline
         $\theta^F$ & $1.0$\\
         \hline
         $v_0^F$ & $1.0$\\
         \hline
    \end{tabular}
    \caption{Fixed values of $\mathbf{p}^F$ for the calibration of the micro model.}\label{tab:fixed_values_micro}
\end{table}
The reason to fix these parameters is that the leverage function overruns their effect. Moreover, we choose $\kappa^F$ and $\theta^F$ so that the Feller condition is not  broken, thus avoiding numerical issues. From a more practical perspective, the impact of such parameters is already tested in \cite{commodity_indices}. \\

For the macro model things are different. Although one could follow the same arguments and fix the parameters controlling the level, we have to recall that we are doing the calibration of the stochastic parameters by calibrating a pure stochastic model. For this reason, all parameters have a potential impact. However, we will also fix the value of $\overline{\kappa}^I$ to avoid breaking the Feller condition, as indicated in Table \ref{tab:fixed_values_macro}.
\begin{table}[H]
    \centering
    \begin{tabular}{|c|c|}
        \hline
         $\bm{\overline{p}}^I$ & \textbf{Value}  \\
         \hline
         $\overline{\kappa}^I$ & $1.0$\\
         \hline
    \end{tabular}
    \caption{Fixed values of $\mathbf{\overline{p}}^I$ for the calibration of the macro model.}\label{tab:fixed_values_macro}
\end{table}
After the previous considerations, the calibration of both models yields the results in Table \ref{tab:calibrated_values_micro_macro}.
\begin{table}[H]
    \centering
    \begin{tabular}{|c|c|c|}
        \hline
         $\bm{p}^F$ & \textbf{Seed} & \textbf{Calibrated Value}  \\
         \hline
         $a$ & $0.3$ & $0.338619$\\
         \hline
         $\beta$ & $0.1$ & $0.172338$\\
         \hline
         $\chi$ & $0.03$ & $1.4$\\
         \hline
         $\rho^{F,v}$ & $-0.2$ & $0.40985$\\
         \hline
    \end{tabular}
    \quad
    \begin{tabular}{|c|c|c|}
        \hline
         $\overline{\bm{p}}^I$ & \textbf{Seed} & \textbf{Calibrated Value}  \\
         \hline
         $\overline{\theta}^I$ & $0.09$ & $0.069918$\\
         \hline
         $\overline{\chi}^I$ & $0.03$ & $0.01277$\\
         \hline
         $\overline{v^I_0}^I$ & $0.09$ & $0.0637628$\\
         \hline
         $\overline{\rho}^{I,v}$ & $-0.2$ & $1.0$\\
         \hline
    \end{tabular}
    \caption{Calibrated values of $\mathbf{p}^F$ and $\overline{\bm p}^I$ against the mean of the plain vanillas on the S\&P GSCI Crude Oil ER index quoted on the 30 November 2019 and on 31 December 2019. For this calibration the hybrid global-local procedure described in Section \ref{sec:calibration} has been used.}\label{tab:calibrated_values_micro_macro}
\end{table}

\subsection{Pricing of some path dependent contracts}
So far we have performed a calibration of both the micro and the macro model. Now, we want to compare the results given by both models when pricing path-dependent options.  However, it is difficult to directly compare the difference in prices between the micro and the macro models. The standard procedure among traders to determine whether the difference is relevant consists in comparing the change in prices among models with the difference in prices that a change of market volatility of $1\%$ would produce in the target contract. More specifically, for the micro model this means shifting the plain vanilla volatilities on the futures $\bm{\sigma}^F_{\text{micro}}$ by $1\%$ and recomputing the leverage of the micro model accordingly:
\begin{equation}
    \begin{aligned}
        & \Delta V_{\text{micro}} = V_{\text{micro}}\left(\bm{\sigma}^F_{\rm market}+1\%\right)-V_{\text{micro}}(\bm{\sigma}^F_{\text{market}})\ .
    \end{aligned}
\end{equation}
For the macro model we proceed in a different way. Since we calibrated the macro model to match the marginal probabilities of the micro one, instead of shifting the market volatilities $\bm{\sigma}^I_{\rm market}$ by one percent, one should shift the micro volatilities $\bm{\sigma}^I_{\rm micro}$ by one percent. Nevertheless, to keep things more consistent, we recalibrate the macro model on the implied volatilities of the index produced by the shifted micro model $\bm{\sigma}^I_{\rm micro}(\bm{\sigma}^F_{\rm market}+ 1\%)$. In this way, both the micro and macro model retain the same marginal probabilities while, at the same time, a shift of $1\%$ in the implied volatilities on the futures approximately produces a shift of $1\%$ in the implied volatilities on the index:
\begin{equation}
    \begin{aligned}
        & \Delta V_{\text{macro}} = V_{\text{macro}}(\bm{\sigma}^I_{\text{micro}}(\bm{\sigma}^F_{\text{market}}+1\%))-V_{\text{macro}}(\bm{\sigma}^I_{\text{micro}}(\bm{\sigma}^F_{\text{market}}))\ .\\
    \end{aligned}
\end{equation}
As a rule of thumbs, the traders consider that, when the model difference is $0.5$ times greater than the difference produced by a change of $1\%$ in the surface of implied volatilities we can say that the difference between models is relevant, i.e.,
\begin{equation}
    \dfrac{V_{\rm micro} - V_{\rm macro}}{\Delta V} \geq 0.5 \longrightarrow V_{\rm micro} \neq V_{\rm macro}.
\end{equation}\label{eq:difference_prices_rule}

We divided the experiments in three sections. In Section \ref{sec:numerical_autocallable} we price autocallable products with three different coupon structures. Next, in Section \ref{sec:numerical_athena} we price a product that is actively traded in the market: the Athena Jet On S\&P GSCI Crude Oil Index Excess Return (see \cite{athena}). Finally, in Section \ref{sec:knock-in} we price a daily barrier knock-in.

All experiments are performed with a confidence level of 95\%.
\subsubsection{Autocallable contracts}\label{sec:numerical_autocallable}
An autocallable is usually structured as a note paying coupons $\{\gamma_1,...,\gamma_{M_A}\}$ on the observation dates $\{T_1,...,T_{M_A}\}$. The contract can be terminated before its expiry on $T_{M_A}$ if a market event occurs. In case of early
termination on $T_i$ a rebate $\beta_i$
is paid along with the capital redemption $\phi_i$. On the last date the whole capital or a part of it $\phi_{M_A}$ is returned along with an optional payoff $\beta_{M_A}$. If we consider for ease of exposition only autocallables on a single underlying
asset, we can describe the market event as the first time the asset price $S_t$ is above a predefined level $H_i$. Thus, the price of an autocallable is
given by
\begin{equation}
    V_t := \mathbb{E}_t\left[\sum_{i = 1}^{M_A}\gamma_i J_iD(t,T_i)+J_{i-1}(1-J_i)(\beta_i+\phi_i)D(t,T_i)\right],
\end{equation}
where the survival indicator $J_i$
is defined for $i\in\{1,...,M_A\}$ as
\begin{equation}
    J_i = \mathds{1}_{\{i<M_A\}}\min_{j\in\{1,...,i\}}\mathds{1}_{\{S_{T_j}<H_j\}},
\end{equation}
where we set $J_0 := 1$ since we assume that at inception the product is alive,
while $H_j$is the barrier level at time $T_j$. We notice that on the last payment
date the autocallable must be terminated, so that on the last date the indicator
is always equal to zero.
The rebate payments are usually equal to the corresponding coupon,
namely
\begin{equation}
    \beta_i = \gamma_i.
\end{equation}
On the other hand, the coupon payments $\gamma_i$ may be defined in different ways.
Here, we briefly describe the most common ones:
\begin{itemize}
    \item Bullet coupons: In this case, the same quantity of money is paid on each date $T_i$, namely
    \begin{equation*}
        \gamma_i^{\text{bullet}} := \overline{\gamma}.
    \end{equation*}
    \item Digital coupons: In this case, a digital option is paid on each date $T_i$, namely
    \begin{equation*}        \gamma_i^{\text{digital}}:=\overline{\gamma}\mathds{1}_{\{S_{T_i}>K_i\}},
    \end{equation*}
where $K_i$ is the digital option strike for the coupon paying at date $T_i$.
    \item Snowball coupons: In this case, a digital option with strike $K$ is paid on each date $T_i$. Moreover, if the option is triggered the amount paid is proportional to the number of coupons not paid since the last payment, namely
    \begin{equation*}        \gamma_i^{\text{snowball}}:=N_i\overline{\gamma}\mathds{1}_{\{S_{T_i}>K_i\}},
    \end{equation*}
    where the proportionality constant is defined as
    \begin{equation*}
        N_i := i-\sup \{0,j\in\{1,...,i-1\} : S_{T_j}>K_j\}.
    \end{equation*}
\end{itemize}
The capital redemptions are only partial to allow to offer higher coupons. Indeed, they are usually defined as
\begin{equation}
    \phi_i := \mathds{1}_{\{i<M_A\}}+\mathds{1}_{\{i=M_A\}}\left(\mathds{1}_{\{S_{T_{M_A}}\geq H_{M_A}\}}+\mathds{1}_{\{S_{T_{M_A}}<H_{M_A}\}}\dfrac{S_{T_{M_A}}}{H_{M_A}}\right),
\end{equation}
so that the capital is wholly redeemed if an early termination is triggered or if the contract is above the prespecified level $H_{M_A}$ on the last date and it is partially redeemed if it is below $H_{M_A}$ on the last date.\\

For the experiments we specifically choose an autocallable that expires nine months after the reference date and can be executed monthly. The barriers and strikes on each callable date are shown in Table \ref{tab:barriers_autocallables}. The barriers have been chosen in such a way that the probability of the contract being executed at any callabillity date is spread throughout the whole callability period. In this way, we try to emphasize the path-dependent properties of the contract, hence the differences among models. 
\begin{table}[H]
\centering
\begin{tabular}{ |c|c|c|} 
 \cline{2-3}
\multicolumn{1}{c|}{} & \cellcolor{g} $\bm{H_i}$ & \cellcolor{g} $\bm{K_i}$\\ 
 \hline
 \cellcolor{g} 2020-01-17 & 1.1 & 1.0\\
 \hline
 \cellcolor{g} 2020-02-17 & 1.1 & 1.0\\
 \hline
  \cellcolor{g} 2020-03-17 & 1.075 & 0.975\\
 \hline
  \cellcolor{g} 2020-04-17 & 1.075 & 0.95\\
 \hline
  \cellcolor{g} 2020-05-19 & 1.075 & 0.925\\
 \hline
  \cellcolor{g} 2020-06-18 & 1.025 & 0.875\\
 \hline
  \cellcolor{g} 2020-07-17 & 0.95 & 0.775\\
 \hline
  \cellcolor{g} 2020-08-17 & 0.85 & 0.675\\
 \hline
  \cellcolor{g} 2020-09-16 & 0.7 & 0.5\\
 \hline
\end{tabular}
    \caption{Barriers for the autocallable contracts.}\label{tab:barriers_autocallables}
\end{table}

With this structure we compute the prices of the product with the three aforementioned coupons for the micro and macro models. The prices and the normalised differences in prices for the micro and macro models are shown in Table \ref{tab:autocallable_prices}. As we can see from the table, both models yield very similar results in terms of prices. Furthermore, the rule of thumbs from Equation \eqref{eq:difference_prices_rule} indicates that we cannot consider that both models are significantly different. 

\begin{table}[H]
\centering
\begin{tabular}{ |c|c|c|c| } 
 \cline{2-4}
\multicolumn{1}{c|}{} & \cellcolor{g} Bullet & 
 \cellcolor{g} Snowball & \cellcolor{g} Digital\\ 
 \hline
 \cellcolor{g} Coupon & 0.005 & 0.005 & 0.005\\
 \hline
 \cellcolor{g} $\bm{V_{\text{macro}}}$ & $1.00996 \pm 0.00018$ & $1.00931 \pm 0.00019$ & $0.997769 \pm 0.00020$\\
 \hline
 \cellcolor{g} $\bm{V_{\text{micro}}}$ & $1.00988 \pm 0.00019$ & $1.0091 \pm 0.00021$ & $0.99769 \pm 0.00021$\\
 \hline
\cellcolor{g} $\bm{\Delta V_{\text{macro}}}$ & $-0.002183 \pm 0.000021$ & $-0.0023147 \pm 0.000023$ & $-0.002329 \pm 0.000022$\\
 \hline
\cellcolor{g} $\bm{\Delta V_{\text{micro}}}$ & $-0.002246 \pm 0.000024$ & $-0.002399 \pm 0.000027$ & $-0.002495 \pm 0.000025$\\
 \hline
 \cellcolor{g} $\dfrac{V_{\text{micro}}-V_{\text{macro}}}{\Delta V_{\text{macro}}}$ & $0.034 \pm 0.121$ & $0.099 \pm 0.122$ & $0.0020 \pm 0.1244$\\[2ex]
 \hline
  \cellcolor{g} $\dfrac{V_{\text{micro}}-V_{\text{macro}}}{\Delta V_{\text{micro}}}$ & $0.033 \pm 0.117$ & $0.095 \pm 0.117$ & $0.001915 \pm 0.1210$\\[2ex]
 \hline
\end{tabular}
\caption{Prices of autocallable contracts with bullet, snowball and digital coupons for the micro and macro models.}\label{tab:autocallable_prices}
\end{table}

\subsubsection{Athena Jet on S\&P GSCI Crude Oil Index Excess Return Certificate}\label{sec:numerical_athena}
In this section we consider a real contract, the Athena Jet on S\&P GSCI Crude Oil Index Excess Return Certificate (see \cite{athena}). This contract has a maximum duration of one year but can expire after the first six months if the price of the S\&P GSCI Crude Oil Index Excess Return index is higher than or equal to its initial value. In that case it returns the nominal value of $1$ euro and pays a $5\%$ premium. Otherwise, the investor receives a premium of
$2.5\%$. At maturity (1 year) the certificate returns $1$ euro and pays a premium equal to 1.5 times the performance of the underlying if the S\&P GSCI Crude Oil Index Excess Return index quotes at a value greater than or equal to its initial value.
If the S\&P GSCI Crude Oil Index Excess Return index quotes between its initial value and a barrier, set at 70\% of the initial value, the certificate protects the invested capital by returning $1$ euro. If instead the S\&P GSCI Crude Oil Index Excess Return quotes below the Barrier, the investor receives an amount commensurate with the performance of the underlying index (with consequent loss on invested capital).\\

The results for this contract are shown in Table \ref{tab:athena_prices}. As in the previous section, we observe that both models produce similar results in terms of prices and normalised differences. However, for this contract the differences become slightly more significant than in Table \ref{tab:autocallable_prices} to the point where we cannot completely discard the possibility that both models differ.

\begin{table}
\centering
\begin{tabular}{ |c|c| } 
 \cline{2-2}
\multicolumn{1}{c|}{} & \cellcolor{g} Athena Jet on S\&P Certificate \\ 
 \hline
 \cellcolor{g} $\bm{V_{\text{macro}}}$ & $1.00207 \pm 0.00048$\\
 \hline
 \cellcolor{g} $\bm{V_{\text{micro}}}$ & $1.00348 \pm 0.00047$\\
 \hline
\cellcolor{g} $\bm{\Delta V_{\text{macro}}}$ & $-0.003718 \pm 0.000091$\\
 \hline
\cellcolor{g} $\bm{\Delta V_{\text{micro}}}$ & $-0.003529 \pm 0.000047$\\
 \hline
 \cellcolor{g} $\dfrac{V_{\text{micro}}-V_{\text{macro}}}{\Delta V_{\text{macro}}}$ & $-0.38 \pm 0.18$\\[2ex]
 \hline
  \cellcolor{g} $\dfrac{V_{\text{micro}}-V_{\text{macro}}}{\Delta V_{\text{micro}}}$
  & $-0.40 \pm 0.19$\\[2ex]
 \hline
\end{tabular}
\caption{Prices of the Athena Jet Su on the S\&P GSCI Crude Oil Index Excess Return for the micro and macro models.}\label{tab:athena_prices}
\end{table}

\subsubsection{Daily Knock In Option}\label{sec:knock-in}
In order to wrap up this section, we check the difference in prices between micro and macro models on a contract that is heavily path dependent. For this purpose we have chosen a daily knock-in option.
\begin{table}
\centering
\begin{tabular}{ |c|c| } 
 \cline{2-2}
\multicolumn{1}{c|}{} & \cellcolor{g} One year daily knock in \\ 
 \hline
 \cellcolor{g} $\bm{V_{\text{macro}}}$ & $0.03656 \pm 0.00026$\\
 \hline
 \cellcolor{g} $\bm{V_{\text{micro}}}$ & $0.03828 \pm 0.00029$\\
 \hline
\cellcolor{g} $\bm{\Delta V_{\text{macro}}}$ & $0.003305 \pm 0.000021$\\
 \hline
\cellcolor{g} $\bm{\Delta V_{\text{micro}}}$ & $0.003425 \pm 0.000027$\\
 \hline
 \cellcolor{g} $\dfrac{V_{\text{micro}}-V_{\text{macro}}}{\Delta V_{\text{macro}}}$ & $0.52 \pm 0.12$\\[2ex]
 \hline
  \cellcolor{g} $\dfrac{V_{\text{micro}}-V_{\text{macro}}}{\Delta V_{\text{micro}}}$
  & $0.50 \pm 0.11$\\[2ex]
 \hline
\end{tabular}
\caption{Prices of a daily knock-in option for the micro and macro models.}\label{tab:knock_in_prices}
\end{table}
Table \ref{tab:knock_in_prices} shows a difference in prices between both models that is closer to being relevant.
\subsection{Sensitivities}\label{sec:numerical_sensitivities}
In this section we explore the effect that both models have in the sensitivities of the Athena Jet On S\&P Crude Oil Index Excess Return. This is interesting because the sensitivities for both the micro and the macro model refer to different quantities. More specifically, in Section \ref{sec:delta} we compute the sensitivity of the micro and the macro model with respect to a change in the level of the future and the index, respectively. Next, in Section \ref{sec:vega} we compute the sensitivity of the micro and the macro model with respect to a change in the level of implied volatilities on the futures and the index, respectively. 
\subsubsection{Delta}\label{sec:delta}
The derivative of the Athena Jet price with respect to the underlying level represents a different thing for the micro and the macro model. 

For the micro model, it represents the change in price with respect to a change in the value of the underlying futures curve. In the case of the first and second futures, its change also produces a change in the value of the index. In order to approximate the derivative we simply shift the values and redo the simulation, so that
\begin{equation}\label{eq:delta_F}
    \Delta^F_i := \dfrac{V_{\text{micro}}((1+10^{-7})\, F_t(T_i))-V_{\text{micro}}(F_t(T_i))}{10^{-7}F_t(T_i)}.
\end{equation}
In the macro model, the delta refers to a change in the level of the index:
\begin{equation}\label{eq:delta_index}
\begin{aligned}
    &\Delta^I_{\text{macro}} := \dfrac{V_{\text{macro}}((1+10^-7)\, I_t)-V_{\text{macro}}(I_t)}{10^{-7}I_t}.\\
\end{aligned}
\end{equation}
\begin{table}[H]
\centering
    \begin{tabular}{|c|c|}
        \hline
        \cellcolor{g}$\bm{\Delta^F_1}$ & \cellcolor{g}$\bm{\Delta^I_{\rm macro}}$ \\
        \hline
        $0.0046046 \pm 0.000029$ & $0.2848 \pm 0.0017$\\
        \hline
    \end{tabular}
    \caption{Delta values for the micro and the macro computed according to Equations \eqref{eq:delta_F} and \eqref{eq:delta_index} respectively.}\label{tab:delta_micro_macro}
\end{table}
When we compute the deltas following Equations \eqref{eq:delta_F} and \eqref{eq:delta_index} we obtain the following the results in Table \ref{tab:delta_micro_macro}. We only show the derivative with respect to the first future because the deltas for the rest of the futures are negligible, actually they are of order $\mathcal{O}(10^{-6})$ or less.\\

It is not surprising that the deltas from both models are different since we are calculating different things. Although this difference might not seem relevant it becomes very important in practice. The reason for this is that it is much easier to trade with futures than with the index, what potentially makes the macro model impractical. For this reason, we propose to define the following quantity in the micro model:
\begin{equation}\label{eq:delta_micro_index}
    \Delta^I_{\rm micro} := \Delta^F_1 \cdot \dfrac{1}{\dfrac{\partial I_t}{\partial F_t(T_1)}}\,.
\end{equation}
The new delta for the micro model defined in Equation \eqref{eq:delta_micro_index} is the result of applying the chain rule to relate the impact that a change on the first future or a change of the index would have in the prices of our target contract. When the deltas are computed in that way the results are shown in Table \ref{tab:delta_micro_macro_index}. In this case, note that the deltas from both the micro and macro models are very similar. Since all the deltas with respect to futures are not meaningful except for the one on the first future and the fact that we have a deterministic relationship between the delta on the first future and the delta on the index, we can conclude that the macro model can be used to compute the delta.
\begin{table}
\centering
    \begin{tabular}{|c|c|}
        \hline
        \cellcolor{g}$\bm{\Delta^I_{\rm micro}}$ & \cellcolor{g}$\bm{\Delta^I_{\rm macro}}$ \\
        \hline
        $0.2738 \pm 0.0017$ & $0.2848 \pm 0.0017$\\
        \hline
    \end{tabular}
    \caption{Delta values for the micro and the macro model computed according to Equations \eqref{eq:delta_micro_index} and \eqref{eq:delta_index} respectively.}\label{tab:delta_micro_macro_index}
\end{table}

\subsubsection{Vega}\label{sec:vega}
In order to complete the study we want to also compare the vegas in the micro and macro models.
The vegas in the micro model are computed by shifting $1\%$ the smile of the plain vanillas on the futures for a fixed maturity (here we denote by $\bm{K}$ the vector of strikes for that specific maturity): 
\begin{equation}
    \mathcal{V}^F_i = \dfrac{V_{\text{micro}}(\sigma_{\text{market}}^{F}(T_i,\bm{K})+0.01)-V_{\text{micro}}(\sigma_{\text{market}}^{F}(T_i,\bm{K}))}{0.01}
\end{equation}
More precisely, we select a futures contract and we only shift the plain vanillas on that specific future. This requires a recalibration of the leverage function. After the recalibration we obtain the results in Table \ref{tab:vega_micro}. We can see that vega is small except for the futures that are close to the autocallability period. However, it may seem counter intuitive that the same effect does not happen when we approach the final payoff. The explanation is that when we change the probability of crossing the barrier in the middle of the contract, that change affects the number of paths that come to the end of the contract. This effect is much greater to that of simply changing the amount that you receive at the end.\\ \\
\begin{table}
\centering
    \begin{tabular}{|c|c|}
        \hline
        \cellcolor{g}$\bm{T_i}$ & \cellcolor{g}$\bm{\mathcal{V}^F_i}$ \\
        \hline
        \cellcolor{g}2020-01-21 & $-0.003 \pm 0.034$\\
        \hline
        \cellcolor{g}2020-02-20 & $-0.027 \pm 0.051$\\
        \hline
        \cellcolor{g}2020-03-20 & $-0.029 \pm 0.063$\\
        \hline
        \cellcolor{g}2020-04-21 & $-0.024 \pm 0.072$\\
        \hline
        \cellcolor{g}2020-05-19 & $-0.144 \pm 0.068$\\
        \hline
        \cellcolor{g}2020-06-22 & $-0.315 \pm 0.065$\\
        \hline
        \cellcolor{g}2020-07-21 & $-0.038 \pm 0.051$\\
        \hline
        \cellcolor{g}2020-08-20 & $-0.053 \pm 0.059$\\
        \hline
        \cellcolor{g}2020-09-22 & $-0.024 \pm 0.052$\\
        \hline
        \cellcolor{g}2020-10-20 & $-0.003 \pm 0.060$\\
        \hline
        \cellcolor{g}2020-11-20 & $-0.024 \pm 0.041$\\
        \hline
        \cellcolor{g}2020-12-21 & $-0.039 \pm 0.026$\\
        \hline
    \end{tabular}
    \caption{Vega values for the micro model.}\label{tab:vega_micro}
\end{table}

 At this point, it is difficult to find an equivalent for the macro model, since there is no direct mapping from the futures to the index. In any case, we suggest a similar procedure for the index, we fix a maturity and shift the market prices in terms of Black-Scholes volatilities by 1\%, \emph{i.e.},
\begin{equation}
    \mathcal{V}^I_i = \dfrac{V_{\text{macro}}(\sigma^I_{\text{micro}}(T^I_i,\bm{K^I})+0.01)-V_{\text{macro}}(\sigma^I_{\text{micro}}(T^I_i,\bm{K^I}))}{0.01}
\end{equation}\label{eq:vegas_macro}
The ``vegas'' $\mathcal{V}^I_i$ calculated with \eqref{eq:vegas_macro} are shown in Table \ref{tab:vega_macro}.
\begin{table}
\centering
    \begin{tabular}{|c|c|}
        \hline
        \cellcolor{g}$\bm{T_i}$ & \cellcolor{g}$\bm{\mathcal{V}^I_i}$ \\
        \hline
        \cellcolor{g}2020-01-16 & $0.022 \pm 0.036$\\
        \hline
        \cellcolor{g}2020-02-17 & $-0.009 \pm 0.060$\\
        \hline
        \cellcolor{g}2020-03-16 & $0.013 \pm 0.059$\\
        \hline
        \cellcolor{g}2020-06-16 & $-0.269 \pm 0.036$\\
        \hline
        \cellcolor{g}2020-12-16 & $-0.030 \pm 0.020$\\
        \hline
    \end{tabular}
    \caption{Vega values for the macro model.}\label{tab:vega_macro}
\end{table}
Both the results of Tables \ref{tab:vega_macro} and \ref{tab:vega_micro} are difficult to compare. The main point in common is that the vegas are greater when approaching the callability period and that both have the same sign. Apart from that, we didn't find any simple relationship between one and the other.
We recall that options on the index are far less liquid than options on futures, thus making the micro model the best solution for hedging.
\subsection{Computer implementation details}
	%%%%%%%%%%%%%%%%%%%%%%%%%%%%%%%%%%%%%%%%%%%%%%%%%%%
	% Computer implementation details
	%%%%%%%%%%%%%%%%%%%%%%%%%%%%%%%%%%%%%%%%%%%%%%%%%%
	Concerning the hardware configuration, all tests have been performed in a Ubuntu server running over a virtualization layer (VMware) with 8 GB of RAM, 64 CPU cores (Intel(R) Xeon(R) CPU E5-2650 v4 at 2.2GHz for a total of 64 logical threads).
	
	On the software side, we have done the implementation in the \textit{C}{\tiny++} programming language. The GNU $C${\tiny++} compiler has been used.

 \section{Conclusions and further research}\label{sec:conclusions}
In this paper we have presented two stochastic local volatility models for the pricing of derivative contracts on commodity indices:
\begin{itemize}
    \item A macro model that directly captures the dynamics of the index.
    \item A micro model that captures the dynamics of the underlying futures curve.
\end{itemize}
Both models have been calibrated to market quotes by means of a hybrid global-local optimization algorithm. 

The macro model produces results that are consistent with the more complex models that capture the behavior of the micro structure. In addition, it is very fast to execute and recalibrate since it requires the simulation of a single asset and the calibration of the stochastic parameters is based on a simple stochastic volatility model. However, it is not able to recover all the sensitivities that are needed for hedging and its calibration depends on data on the index, information that is scarce. Typically, the set of plain vanillas on the index is obtained from a consensus. The consensus is built from the data provided by different financial institutions on a monthly basis. Those financial institutions providing data that differs significantly from the consensus are expelled from the consensus data source. If we try to use it on a daily basis we will find that we either recalibrate the model once a month or we are forced to use another model that provides us a more stable stream of information.

The micro model is very robust. It produces prices for the derivatives on the index that are consistent with the models on the index. Its sensitivities refer to products that are actively traded in the market what makes it better for hedging and trading purposes. Furthermore, it can be recalibrated on a daily basis, the stochastic parameters can be kept fixed and we can adjust the leverage function by recalibrating it to plain vanillas on the futures curve. Nevertheless, it is a very heavy model. The simulation of a contract depending on the index that expires one year into the future required the simulation of approximately $12$ different SLV processes and the situation becomes worse as we go further into the future.

In view of the previous arguments, each model has its strengths and weaknesses and each of them can be effective on a different context.

Among the possible future research lines, we specifically mention two. On the one hand, we could price other indices, presumably  multi-commodity ones. On the other hand, we could explore alternative models for the stochastic volatility. A possibility comes from extending the Heston model to admit powers of the stochastic volatility in the diffusion term of the stochastic volatility dynamics. This extension would be motivated by the fact that none of the parameters $(\kappa,\theta,v_0)$ of the stochastic volatility affected the prices of the plain vanillas on the index. This suggests that a further modifications of the drift term will not have any impact on the prices of plain vanillas. However, the diffusion term had an impact. Therefore, by considering powers of the volatility in the diffusion term of the stochastic volatility dynamics we could enhance the calibration of the plain vanilla options on the index.
\section*{Acknowledgements} 
A.P. Manzano-Herrero and C. V\'azquez acknowledge the support from CITIC, as a center accredited for excellence within the Galician University System and a member of the CIGUS Network, receives subsidies from the Department of Education, Science, Universities, and Vocational Training of the Xunta de Galicia. Additionally, it is co-financed by the EU through the FEDER Galicia 2021-27 operational program (Ref. ED431G 2023/01). Also, both authors acknowledge the funding from Xunta de Galicia through the grant ED431C 2022/47, as well as the funding from Spanish Ministry of Science and Innovation with the grant PID2022-141058OB-I00.
\printbibliography

@article{nastasi2020smile,
  title = "Smile modeling in commodity markets",
  author = "Nastasi, E. and Pallavicini, A. and Sartorelli, G.",
  journal = "International Journal of Theoretical and Applied Finance",
  volume = 23,
  number = 3,
  pages = "1--28",
  year = "2020",
  publisher = "World Scientific"
}

@book{wilmott1995mathematics, place={Cambridge}, title={The Mathematics of Financial Derivatives: A Student Introduction}, publisher={Cambridge University Press}, author={Wilmott, Paul and Howison, Sam and Dewynne, Jeff}, year={1995}}

@article{FERREIROFERREIRO2020467,
title = "A new calibration of the {H}eston {S}tochastic {L}ocal {V}olatility Model and its parallel implementation on {GPUs}",
journal = "Mathematics and Computers in Simulation",
volume = 177,
pages = "467--486",
year = 2020,
author = "Ferreiro-Ferreiro, Ana M. and Garc\'ia Rodr\'iguez, José Antonio A. and Souto, Luis and V\'azquez, Carlos",
}

@article{aarts1985statistical,
  title = "Statistical cooling: A general approach to combinatorial optimization problems",
  author = "Aarts, E. H. L. and Van Laarhoven, P. J. M.",
  journal = "Philips Journal of Research",
  volume = 40,
  number = 4,
  pages = "193--226",
  year = 1985,
  publisher = "Philips Research Laboratories"
}

@article{storn1997differential,
  title = "Differential evolution -- a simple and efficient heuristic for global optimization over continuous spaces",
  author = "Storn, R. and Price, K.",
  journal = "Journal of Global Optimization",
  volume = 11,
  number = 4,
  pages = "341--359",
  year = 1997,
}

@INPROCEEDINGS{kennedy1995particle,
  author={Kennedy, J. and Eberhart, R.},
  booktitle={Proceedings of ICNN'95 - International Conference on Neural Networks}, 
  title={Particle swarm optimization}, 
  year={1995},
  volume={4},
  number={},
  pages={1942-1948 vol.4},
  doi={10.1109/ICNN.1995.488968}}

@article{lord2010comparison,
  title = "A comparison of biased simulation schemes for stochastic volatility models",
  author = "Lord, R. and Koekkoek, R. and Van Dijk, D.",
  journal = "Quantitative Finance",
  volume = 10,
  number = 2,
  pages = "177--194",
  year = 2010,
}

@article{guyon2012being,
  title = "Being particular about calibration",
  author = "Guyon, Julien and Henry-Labord{\`e}re, Pierre",
  journal = "Risk Magazine",
  volume = 25,
  number = 1,
  pages = "88--93",
  year = 2012,
  publisher = "Incisive Media Limited"
}

@article{gyongy1986mimicking,
  title = "Mimicking the one-dimensional marginal distributions of processes having an {It{\^o}} differential",
  author = "Gy{\"o}ngy, I.",
  journal = "Probability Theory and Related Fields",
  volume = 71,
  number = 4,
  pages = "501--516",
  year = 1986
}

@article{dupire1994pricing,
  title = "Pricing with a Smile",
  author = "Dupire, Bruno",
  journal = "Risk Magazine",
  number = 1,
  pages = "18-20",
  year = 1994
}

@article{Derman1994,
  title = "Riding on a smile",
  author = "Derman, E. and Kani, I.",
  journal = "Risk Magazine",
  number = 7,
  pages = "32-39",
  year = 1994
}

@article{Pilz2011,
  title = "A hybrid commodity and interest rate market model",
  author = "Pilz, K. and Schl{\"o}gl, E.",
  journal = "Quantitative Finance",
  number = 13,
  volume = 4,
  pages = "543-560",
  year = 2011
}

@book{rebonato1999volatility,
  title = "Volatility and correlation: In the pricing of equity, {FX} and interest-rate options",
  author = "Rebonato, Ricardo",
  year = 1999,
  publisher = "John Wiley \& Sons"
}

@article{heston1993closed,
  title = "A closed-form solution for options with stochastic volatility with applications to bond and currency options",
  author = "Heston, S. L.",
  journal = "The Review of Financial Studies",
  volume = 6,
  number = 2,
  pages = "327--343",
  year = 1993
}

@article{said1999pricing,
  title = "Pricing exotics under the smile",
  author = "Said, K.",
  journal = "Risk Magazine",
  volume = 12,
  pages = "72--75",
  year = 1999
}

@article{lipton2002masterclass,
  title = "The vol smile problem",
  author = "Lipton, Alexander",
  journal = "Risk Magazine",
  volume = 15,
  number = 2,
  pages = "61--66",
  year = 2002
}

@article{ren2007calibrating,
  title = "Calibrating and pricing with embedded local volatility models",
  author = "Ren, Yong. and Madan, Dilip and Qian Qian, Michael",
  journal = "Risk Magazine",
  volume = 20,
  number = 9,
  pages = "138--143",
  year = 2007
}

@ARTICLE{DE_variation,

  author={da Silva Santos, Carlos Henrique and Gon\c{c}alves, Marcos Sergio and Hern\'andez-Figueroa, Hugo Enrique},

  journal={IEEE Photonics Technology Letters}, 

  title={Designing Novel Photonic Devices by Bio-Inspired Computing}, 

  year={2010},

  volume={22},

  number={15},

  pages={1177-1179},

  doi={10.1109/LPT.2010.2051222}}

@article{Nelder1965ASM,
  title={A Simplex Method for Function Minimization},
  author={John A. Nelder and Roger Mead},
  journal={The Computer Journal},
  year={1965},
  volume={7},
  number={4},
  pages={308-313}
}

@phdthesis{subplex,
author={T. Rowan},
title={Functional Stability Analysis of Numerical Algorithms},
school={Department of Computer Sciences, University of Texas at Austin},
year={1990}}

@INPROCEEDINGS{EA,

  author={Vikhar, Pradnya A.},

  booktitle={2016 International Conference on Global Trends in Signal Processing, Information Computing and Communication (ICGTSPICC)}, 

  title={Evolutionary algorithms: A critical review and its future prospects}, 

  year={2016},

  volume={},

  number={},

  pages={261-265},

  doi={10.1109/ICGTSPICC.2016.7955308}}

@misc{chiminello,
  title = "Oil goes local",
  author = "Chiminello, F.",
  note = "Talk at Imperial College, London",
  year = 2015
}

@article{two_phase,
title = {Parallel two-phase methods for global optimization on GPU},
journal = {Mathematics and Computers in Simulation},
volume = {156},
pages = {67-90},
year = {2019},
issn = {0378-4754},
doi = {https://doi.org/10.1016/j.matcom.2018.06.005},
url = {https://www.sciencedirect.com/science/article/pii/S0378475418301502},
author = {Ana M. Ferreiro and José Antonio García-Rodríguez and Carlos Vázquez and Eliana Costa {e Silva} and Aldina Correia},
}

@misc{NLopt,
  title = {The {NLopt} nonlinear-optimization package},
  author = {Steven G. Johnson},
  year = {2007},
  howpublished = {\url{https://github.com/stevengj/nlopt}}
}

@online{athena,
title={Certificate Athena Jet Su S\&P GSCI Crude Oil Index Excess Return},
author={BNP Paribas},
url={https://www.borsaitaliana.it/cw-e-certificates/covered-warrant/brochureathenajet.pdf},
journal={Borsa Italiana},
year={2015}
}

@article{autocallable1,
author = {Deng, Geng and Mccann, Craig and Mallett, Joshua},
year = {2011},
month = {08},
pages = {326-340},
title = {Modeling Autocallable Structured Products},
volume = {17},
journal = {Journal of Derivatives \& Hedge Funds},
doi = {10.1057/jdhf.2011.25}
}

@article{autocallable2,
author = {Alm, Thomas and Harrach, Bastian and Harrach, Daphne and Keller, Marco},
year = {2013},
month = {09},
pages = {43-70},
title = {A Monte Carlo pricing algorithm for autocallables that allows for stable differentiation},
volume = {17},
journal = {The Journal of Computational Finance},
doi = {10.21314/JCF.2013.265}
}

@article{autocallable3,
  title={Pricing autocallables under local-stochastic volatility},
  author={Farkas, Walter and Ferrari, Francesco and Ulrych, Urban},
  journal={Swiss Finance Institute Research Paper},
  number={22-71},
  year={2022}
}

@article{commodity_indices,
author = {Alberto Manzano and Emanuele Nastasi and Andrea Pallavicini and Carlos V\'azquez},
title = {Pricing commodity index options},
journal = {Quantitative Finance},
volume = {23},
number = {2},
pages = {297-308},
year  = {2023},
publisher = {Routledge},
doi = {10.1080/14697688.2022.2138775},
URL = {https://doi.org/10.1080/14697688.2022.2138775},
eprint = { https://doi.org/10.1080/14697688.2022.2138775}
}

@article{hagan2002managing,
  title={Managing smile risk},
  author={Hagan, Patrick S and Kumar, Deep and Lesniewski, Andrew S and Woodward, Diana E},
  journal={The Best of Wilmott},
  volume={1},
  pages={249--296},
  year={2002}
}

 \end{document}